\documentclass{emulateapj}
\usepackage{graphicx}

\bibstyle{aas}

\begin{document}

\submitted{To be submitted to ApJ}
\author{Gilbert P. Holder\altaffilmark{1}, 
Kenneth M. Nollett\altaffilmark{2}, 
Alexander van Engelen\altaffilmark{1}}
\altaffiltext{1}{Department of Physics, McGill University, Montreal QC H3A 2T8}
\altaffiltext{2}{Physics Division, Argonne National Laboratory, 9700 S. Cass Ave., Argonne, IL~~60439}

\title{On Possible Variation in the Cosmological Baryon Fraction}

\begin{abstract}

The fraction of matter that is in the form of baryons or dark matter
could have spatial fluctuations in the form of baryon-dark matter
isocurvature fluctuations.  We use big bang nucleosynthesis
calculations compared with observed light element abundances as well
as galaxy cluster gas fractions to constrain cosmological variations
in the baryon fraction.  Light element abundances constrain spatial
variations to be less than 26-27\%, while a sample of ``relaxed''
galaxy clusters shows spatial variations in gas fractions less than
8\%.  Larger spatial variations could cause differential screening of
the primary cosmic microwave background anisotropies, leading to
asymmetries in the fluctuations and ease some tension with the
halo-star $^7\mathrm{Li}$ abundance. Fluctuations within our allowed
bounds can lead to ``$B$-mode'' CMB polarization anisotropies at a
non-negligible level.
\end{abstract}

\section{Introduction}

Is the cosmological fraction of matter that is in the form of baryons
a universal constant?  There is no compelling theoretical model that
would predict it to be observably non-constant (although such an
effect would be fairly easy to ``explain''), but there is little
empirical evidence that it is indeed a constant to high precision.

The cosmological gravitational potential fluctuations are observed to
be at the level of $10^{-5}$, but this measures fluctuations in the
sum of baryonic and non-baryonic mass.  The fraction of matter in the
form of baryons could have much larger long wavelength fluctuations
with little observable imprint in the cosmic microwave background
(CMB). Such a modulation is a baryon-cdm isocurvature mode, and its
imprint on the CMB is a second order effect. On scales larger than the
sound horizon at recombination there is almost no effect, while on
scales that are sub-horizon at last scattering there would be an
additional modulation of the acoustic peaks due to a varying sound
speed. The extra pressure due to the baryons is negligible compared to
the photon pressure.
 
In this paper we investigate other effects of the variation of the
baryon fraction, and we find that large scale modulation is not only
allowed but would actually alleviate some current mild tensions in the
standard cosmology.

In particular, we below investigate constraints from big-bang
nucleosynthesis (BBN) in light of observed light element abundances,
observations of galaxy cluster gas fractions, and modulation of the
optical depth to Thomson scattering of CMB photons. We also
investigate the expected contribution to CMB polarization anisotropy
and discuss possible other probes.

\section{Big bang nucleosynthesis}

The light-nuclide yields from BBN are functions only of the baryon
density $\rho_B$ of the universe under the standard assumptions of
uniform entropy per baryon, only standard-model particles,
neutrino-antineutrino asymmetries not enormously larger than the
corresponding baryon asymmetry, and no late additions of entropy
\citep{schramm98}.  Observations of extragalactic deuterium provide a
particularly tight constraint on $\rho_B$ based on the steep
dependence of the deuterium yield on this parameter.  Using
$\Omega_BH_o^2= 8\pi G \rho_B/3$ to express the constraint in terms of
the fraction $\Omega_B$ of the closure density provided by baryons,
the average extragalactic D/H ($\equiv$ ratio by number of deuterium
to hydrogen) gives $\Omega_Bh^2=0.0213\pm 0.0010$ \citep{pettini08}.
($H_0 =h\times 100\ \mathrm{km\ s^{-1}\,Mpc^{-1}}$ is the Hubble
``constant.'')  CMB anisotropy measurements with the WMAP satellite
now imply $\Omega_Bh^2=0.02273\pm 0.00062$ \citep{dunkley09}, in
agreement with D/H.  In these terms, the baryon fraction of the total
mass density $\rho_M$ is $f_B=\rho_B/\rho_M$.

There is an extensive literature on the effects of inhomogeneities in
the baryon distribution during BBN, going back to at least
\cite{wagoner73} and \cite{epstein75}.  For a review of
baryon-inhomogeneous and other non-standard BBN models up to 1993, see
\cite{malaney93}.  

Baryon-inhomogeneous scenarios of BBN may be divided into three cases,
depending on the typical scale of the fluctuations: scales that are
comparable to particle-diffusion horizons during BBN (such that the
physics of BBN is affected), larger scales that are still small
compared to the scale of our Galaxy (such that observations within our
Galaxy are a blend of many different volumes that individually
underwent nearly homogeneous BBN), and large scales today (such that
cosmological scales sample distinct regions of uniform baryon
density).

In the first case, which dominates the literature on inhomogeneous
BBN, the inhomogeneities are comparable in length scale to the neutron
diffusion length during BBN ($\sim 4\times 10^{-5}$ comoving pc at the
start or $\sim 0.08$ comoving pc at the end of BBN; \cite{kurki97}).
Such short-length-scale inhomogeneities only have strong effects if
the density contrasts are large (ratios of $\sim 10^6$ between high-
and low-density domains), as was suggested in the 1980s on the basis
of a strongly first-order QCD phase transition \citep{witten84}.  A
first-order phase transition of this kind has since fallen out of
favor but is not completely ruled out \citep{boyanovsky06}.  The
corresponding BBN models tend to overproduce lithium and underproduce
deuterium relative to current measurements; for a recent assessment of
constraints, see \cite{lara06}.  This is not the case that we consider
in the present work.

Inhomogeneities longer than the diffusion length result in BBN at each
location occurring independently at different $\rho_B$.  These were
investigated by \cite{epstein75}, and the most elaborate calculations
were performed with a view toward constraining isocurvature baryon
fluctuations as seeds of structure formation
\citep{jedamzik95,copi95,kurki97}.  These authors assumed that any
single observation is a blended sample containing matter drawn from
the full distribution of $\rho_B$; the resulting effect on BBN is a
``smearing out'' of the standard BBN predictions of abundance versus
mean baryon density.  Calculations in the three works last cited
mostly excluded perturbations exceeding the Jeans mass at
recombination (which would not be observable sites of low metallicity
today), with the effect of diminishing the contributions of
high-$\rho_B$ domains.  D/H then becomes much larger than observed and
the primordial $^4$He mass fraction $Y_P$ is reduced relative to the
homogeneous case.

The main constraint on blended inhomogeneities arose from flattening
of the Li/H versus $\rho_B$ curve, which limits mixed inhomogeneities
to $\delta \rho_B/\rho_B\lesssim 1$.  If the Li/H observed in
low-metallicity halo stars is undepleted, the primordial Li/H is
constrained to lie near a minimum of the Li/H versus $\rho_B$ curve of
standard BBN.  Any mixing with regions of $\rho_B$ far from the
minimum increases Li/H unacceptably. (See in particular Figs. 1-3 of
\cite{kurki97} for these results).  The difference of the WMAP- and
D/H-inferred $\rho_B$ from the Li/H minimum obviously complicates this
inference.

The type of inhomogeneity to be addressed here is on length scales
that did not mix during structure formation, so that abundance
measurements are not blends.  This possibility was noted in earlier
work, particularly in light of large differences between the early
claims of extragalactic D/H measurements \citep{kurki97,copi98}.  It
was not extensively pursued because most observational constraints
(all Li/H and $^3$He/H, and all D/H before 1995) were within the
Galaxy and presumed to constitute a single well-mixed sample, and
because the higher claimed D/H values were eventually discounted.

We now examine the implications of the existing light-nuclide
abundance data for unmixed inhomogeneities.  The emphasis in the
earlier literature was on how inhomogeneities affected the BBN upper
bound on $\Omega_B$; we take $\Omega_B$ as a more or less settled
quantity and ask how large fluctuations away from it can be.

\subsection{$^2\mathrm{H}$}

The strongest BBN probe of $\rho_B$ is deuterium, with
$\mathrm{D/H}\propto \rho_B^{-1.6}$.  It also constrains site-to-site
variations, since observations of deuterium believed to be primordial
span a range of redshifts from about 2.5 to 3.6 at varying directions
on the sky; the corresponding length scales are comparable to the
present cosmological horizon.

\begin{figure}
\centerline{
\includegraphics[height=1.9in,width=1.8in]{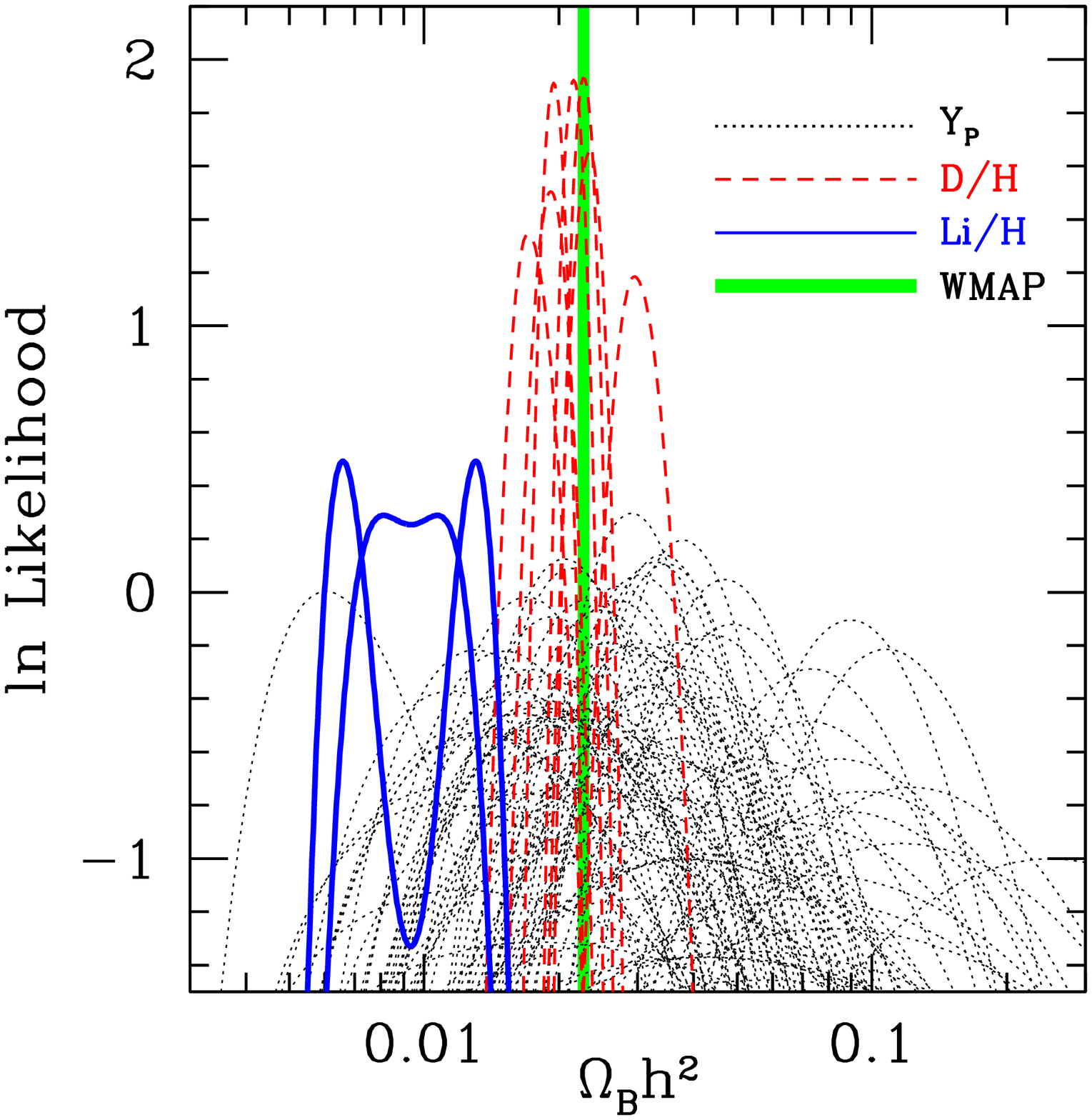}
\includegraphics[trim=30 40 0 0, width=1.8in,height=1.9in]{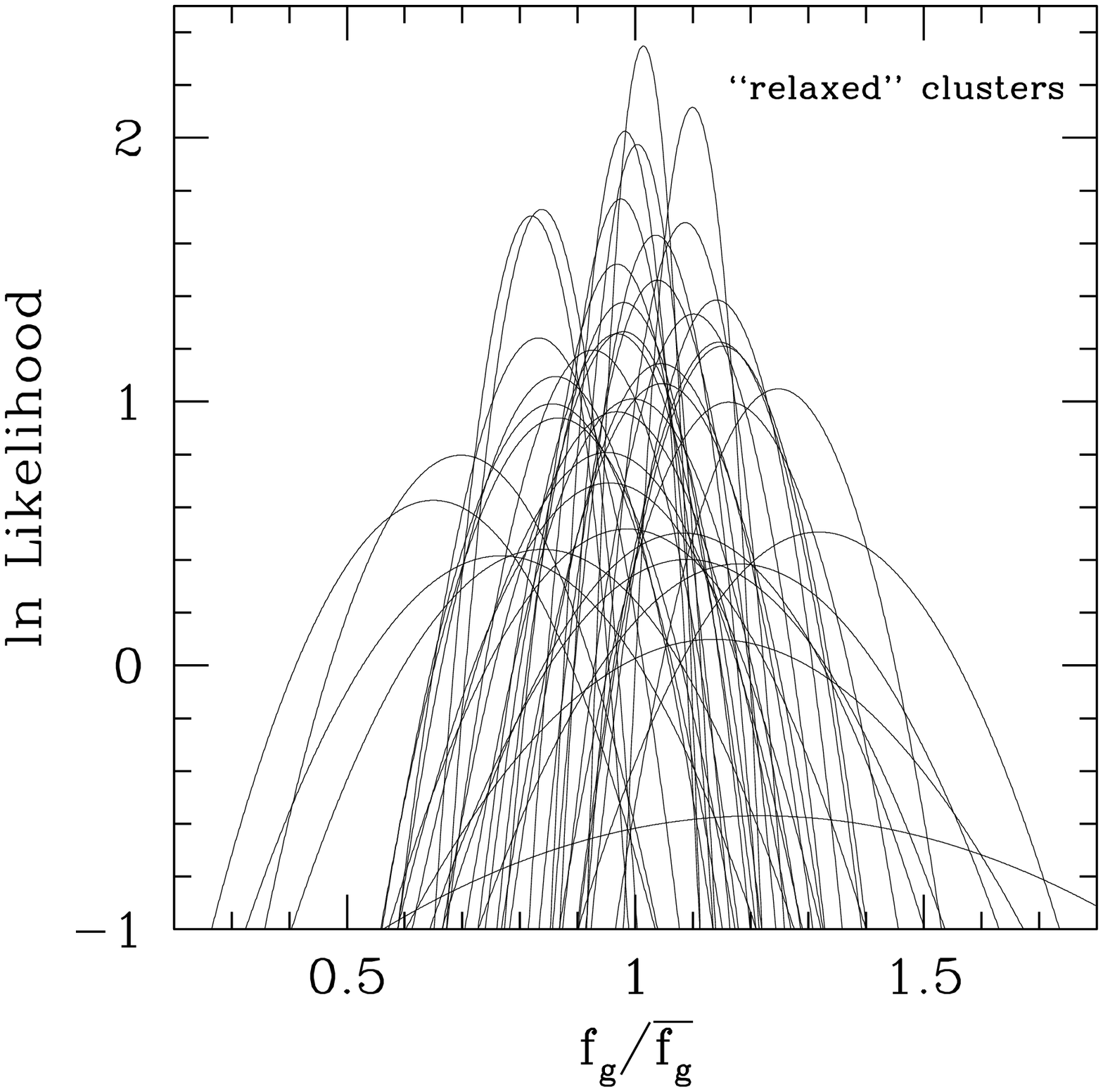} 
}
  \caption{(Left) Likelihoods for $\Omega_B h^2$ based on measured abundances
    in various locations, shown against a band indicating
    the $1\sigma$ interval from WMAP5.   The
    BBN yields of \cite{burles01} were used, assuming Gaussian
    distribution of observational errors only; error estimates for 
    the BBN calculation are not included.
(Right)
Galaxy cluster gas fraction probability distributions relative
to mean of sample, assuming Gaussian probability distributions for
sample of \cite{allen08}.}
\label{fig:bbn-likelihoods}
\label{fig:cluster_probs}
\end{figure}

In the dashed curves of Fig. \ref{fig:bbn-likelihoods}, we show
likelihood as a function of $\Omega_Bh^2$ derived from seven
measurements of D/H in quasar absorption-line systems.  These are the
results tabulated in \cite{kirkman03} [from
  \cite{tytler96,burlestytler98a,burlestytler98b,omeara01,pettini01,kirkman00}]
plus those of \cite{omeara06} and \cite{pettini08}.  Following the
discussion in \cite{kirkman03}, \cite{omeara06}, and \cite{pettini08},
we have omitted from the figure and subsequent analysis the D/H toward
Q0347-3819 \citep{levskakov02} and the absorber at $z=3.256$ towards
PKS1937-1009 \citep{crighton04} (the error on the latter being claimed
to be underestimated).

It is well established in the literature that there is dispersion
within the D/H sample in excess of the reported uncertainties, so that
for example a jackknife estimate of the error on $\log\mathrm{D/H}$ is
about twice that derived by the inverse-variance estimate
\citep{kirkman03,omeara06,pettini08}.  It may simply be that the
errors have been underestimated.  Alternatively, one can accept the
published error estimates and use the data to limit variation in
$\rho_B$ among the locations where deuterium is observed.

\subsection{$^4\mathrm{He}$}

Because the primordial $^4$He mass fraction $Y_P$ depends very weakly
on $\rho_B$, useful constraints require very difficult percent-level
abundance determinations in sites of low metallicity.  Operationally,
this means observations of H{\sc ii} regions in blue compact galaxies
(BCG) or the Magellanic clouds (most recently
\cite{olive97,peimbert00,peimbert07,izotov07}).  Since these
observations are extragalactic, they might be sensitive to large-scale
abundance variations.  The sample of H{\sc ii} regions well-measured
for $Y_P$ extends only to $z\sim 0.05$, or $\sim 200 h^{-1}$ Mpc.
\cite{izotov07} recently added to the available sample 271 blue
compact galaxies identified from the Sloan Digital Sky Survey (SDSS);
most of these are in the same redshift range as the previous sample,
but there are tens of objects in the range 200 to 500 $h^{-1}$ Mpc
($0.05 \lesssim z \lesssim 0.13$).

The inferred $Y_P$ values are known to suffer from systematic
difficulties.  Disagreement among groups using different methods
exceeds their claimed errors, and assessments of the situation are
sometimes very pessimistic \citep{olive04}.  After a recent revision
of the atomic data on which $Y_P$ rests, some researchers find values
in good agreement with WMAP \citep{peimbert07}, while others find
$Y_P$ too high for a comfortable match \citep{izotov07}.

The systematic problems with $Y_P$ should not matter for present
purposes, because spatial variation of $Y_P$ should show up as
dispersion within any uniformly-analyzed data set.  However,
dispersion is not as simple as variation away from the mean $Y_P$,
because truly primordial $Y$ is not measured anywhere.  All H{\sc ii}
regions observed for $Y_P$ have small but nonzero metal content, and
the same stars that made the metals also made $^4$He.  Even though
systems observed for $Y_P$ are chosen for low metallicity, inferred
values of $Y_P$ are almost always based on extrapolation to
$\mathrm{O/H}=0$.  One either applies a $Y$ versus O/H relation
calibrated elsewhere or carries out a linear regression of $Y$ versus
O/H among observed H{\sc ii} regions.  Despite these complications, no
signs of dispersion beyond the error bars have been claimed in the modern
data set.

In our analysis, we place limits on dispersion within the large
``HeBCG'' sample of 93 uniformly-analyzed H{\sc ii} regions from
\cite{izotov07}, with $Y$ corrected to $Y_P$ by subtracting the $Y$
versus O/H regression slope derived in \cite{izotov07}.  We have not
included the additional SDSS H{\sc ii} regions from the same
publication.  These were not subjected to the same data reduction as
the HeBCG sample; they were selected differently; they are overall of
higher metallicity; their regressed $Y_P$ are consistent with the
HeBCG sample at the $2\sigma$ level, but regressions of the two sets
have rather different slopes.  The errors on the SDSS sample are also
large enough that a joint regression of both sets is still dominated
by the smaller HeBCG sample \citep{izotov07}.  The constraints on
$Y_P$ based on each of the 93 H{\sc ii} regions are shown as dotted
curves in Fig. \ref{fig:bbn-likelihoods}.

\subsection{$^7\mathrm{Li}$}

Measurements of primordial lithium are confined to the atmospheres of
metal-poor stars in the Galactic halo.  Since our present interest is
in variations of $\rho_B$ on much larger length scales, all of these
stars together constitute a single sample.

Dispersion of Li/H among the stars of the ``Spite plateau'' -- the
flat region in the graphs of Li/H versus both effective temperature
and metallicity -- has been actively sought ever since the discovery
of the plateau and its identification with the primordial Li/H
\citep{spite82}.  This is because star-to-star variation along the
plateau would be a strong indication that the observed lithium has
been depleted \citep{deliyannis93}, the dispersion having arisen from
varying amounts of depletion.  Some authors claimed dispersion of 0.04
to 0.1 dex \citep{thorburn94,deliyannis93}, but more recent work has
mostly stressed its absence
\citep{ryan99,charbonnel05,asplund06,bonifacio07}.  In the solid
curves of Fig. \ref{fig:bbn-likelihoods}, we show the galactic halo as
a single sample of Li/H, using the values from \cite{asplund06} and
\cite{bonifacio07}.  Primordial Li/H values published over a 30-year
span all agree within the quoted errors.

These observations at face value indicate that our lone sample of
primordial Li/H is a factor of three below the Li/H predicted by the
$\Omega_Bh^2$ from WMAP.  Spatial variations about the WMAP mean would
help to explain this discrepancy if we live in a region of low
$\rho_B$.  However, inspection of Fig. \ref{fig:bbn-likelihoods} shows
that variations in excess of 50\% would be needed for a complete
explanation.  This seems implausible based on constraints derived
below.

\subsection{$^3\mathrm{He}$}

There is no convincingly primordial observed $^3$He/H: all
observations of $^3$He are near solar metallicity, and a strong
increase of $^3$He/H with time is predicted from models of stellar
nucleosynthesis.  Measurements of $^3$He/H in various locations within
the Galaxy \citep{geiss98,gloeckler96,bania02,bania07} fail to see any
evolution of $^3$He/H and are all near the WMAP-inferred primordial
value.  Presumably some mechanism of $^3$He destruction is active, and
its outlines seem clear \citep{hogan95,wasserburg95,charbonnel95}.
Among other things, it causes difficulties for the traditional
$\mathrm{(D+\!^3He)/H}$ constraint applied in the earlier literature
on inhomogeneous BBN.

Since all existing $^3$He/H measurements are within the Galaxy, are
near the WMAP-inferred prediction, and suffer from large uncertainties
in the post-BBN evolution, they do not provide constraints on
variation of $\rho_B$.

\section{Galaxy cluster gas fractions}

Measurements of the galaxy cluster gas fraction can provide strong
constraints on the cosmic baryon fraction, provided the
galaxy clusters adequately sample the cosmic baryon fraction
distribution \citep{white93}. 
Galaxy clusters emit in the X-ray through a combination
of thermal bremsstrahlung and line emission, allowing detailed 
reconstructions of both density and temperature profiles. The electron
density can be directly integrated to obtain the total gas mass, while
the temperature information can be added to obtain an estimate of the
 total mass (assuming hydrostatic equilibrium). 

The sample of \citet{allen08} has gas fraction determinations of 
42 galaxy clusters covering the redshift range $z\sim 0.06-1$. 
This is a large homogeneously analyzed sample of galaxy clusters
that was primarily selected to include objects that appear
to be relatively undisturbed (``relaxed'').

In practice, there are real concerns about the importance of the
assumption of spherical or biaxial symmetry (even triaxial ellipsoidal
symmetry is a strong assumption), deviations from equilibrium,
non-thermal pressure support, non-equilibration of electron and proton
temperatures, and other issues. However, empirically, this method has
been found to be relatively robust when applied to simulated clusters
with many real-world complications \citep{nagai07}, and the
restriction to only include relatively undisturbed clusters should
significantly reduce many of these complications.  In general, it is
found that galaxy cluster gas fractions are below the cosmic mean when
measured well within the virial radius \citep{vikhlinin06, allen08},
but there is little evidence for cluster-to-cluster scatter, as seen
in Figure \ref{fig:cluster_probs}.  \cite{allen08} find
$f_\mathrm{gas} = 0.1105\pm 0.0016$, while the WMAP5 baryon
fraction is $0.17\pm 0.01$.

\newpage
\section{Constraints on Baryon Fraction Variations}

A maximum likelihood fit was done to each data set, varying the 
mean, $m$,  and intrinsic scatter, $s$. The fits were done to the
gas fractions and abundance determinations.
The figure of merit used was
\begin{equation}
-2\ln L = \sum_i \Bigl[ { (d_i - m)^2 \over s^2 + \sigma_i^2 }
  + \ln(s^2+\sigma_i^2)  \Bigr]\,,
\end{equation}
with $d_i$ the data and $\sigma_i$ their errors.

Marginalizing over the mean, an estimate of the likelihood of the
intrinsic scatter was obtained for each data set. For the galaxy
cluster gas fractions, this was translated directly into scatter in
the fractional baryon fraction.  The fits of \citet{burles01},
expanded around the best fit WMAP5 value for $\Omega_B
h^2$\citep{dunkley09}, were used to translate the fitted scatter in
each light-nuclide abundance into scatter in the baryon density.
Linear fits were applied to the BBN yields in the region of interest,
providing:
\begin{equation}
{ \Delta \Omega_B h^2 \over \Omega_B h^2} \sim {\Delta Y_p \over 0.0087}
\end{equation}
for the Helium abundance and
\begin{equation}
{ \Delta \Omega_B h^2 \over \Omega_B h^2} \sim {\Delta [\log_{10}(D/H)] \over 0.69}
\end{equation}
for the Deuterium abundance.

\begin{figure}
\plotone{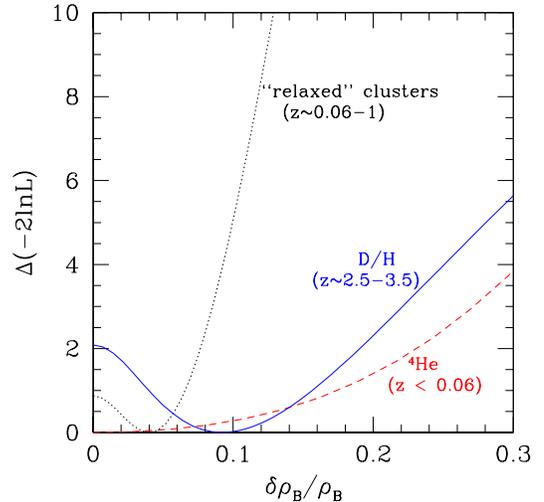}
\caption{Probability distribution of scatter parameter for various
data sets}
\label{fig:chisq}
\end{figure}

Results of the likelihood calculations are shown in Figure
\ref{fig:chisq}.  Upper limits on intrinsic scatter in the baryon
fraction were obtained by integrating the likelihoods starting at
$s=0$ until 95\% of the probability was enclosed. Constraints from
$^4$He and D/H allow 95\% confidence upper limits on intrinsic scatter
of $\sigma(f_B/\bar{f_B})<0.27$ and 0.26, respectively. From galaxy
cluster gas fractions, this limit is only 0.08.  (We note that the
best-fit $s$ for $\log_{10}\mathrm{D/H}$ at $s\sim 0.07$
($\delta\rho_B/\rho_B\sim 0.10$) in Fig. \ref{fig:chisq} is in
agreement with a similar analysis by \cite{pettini08}.)

The scales probed by these measurements vary greatly: the $^4$He
measurements lie at low redshift (less than 0.1), the galaxy clusters
are at intermediate redshift (less than 1), and the D/H constraints are
at $z\sim 2.5-3.5$.

The possible selection effects are also very different. The $^4$He
constraints come from a large sample of nearby low-metallicity
galaxies experiencing star formation today.  The D/H measurements
require a strong signature of D in a QSO absorption spectrum (hence a
large hydrogen column density) and favorable conditions for fitting D
and H column densities (hence a simple velocity structure with minimal
nearby absorption and a bright QSO).

The galaxy cluster sample is selected for being preferentially
``relaxed'' in their appearance. This means appearing undisturbed and
relatively circular. In a universe with varying baryon fractions, such
selection could be problematic: low baryon regions are likely to have
more density fluctuations on small scales \citep{eisenstein98}, and
clusters there are likely to form earlier. If a sample is selected for
clusters that are apparently dynamically more mature (``relaxed''),
this sample could be skewed toward regions of lower baryon fraction,
where clusters formed earlier. A sample of galaxy clusters selected in
this way would thus show less scatter than that of the universe as a
whole, and would be systematically biased toward regions of lower
baryon fraction. Detailed numerical simulations of galaxy cluster
formation will be required to better understand this important
selection effect.

In summary, there is no evidence for cosmological variation in the
baryon fraction, but the amplitude is constrained to be no larger
than roughly 10-30\%.
For comparison, baryon-photon or cdm-photon isocurvature modes have
been constrained to have amplitudes on horizon scales at the level
of $10^{-5}$ or smaller (e.g., \citet{sollom09} and references
therein).

\section{Differential Thomson scattering}

In recent years, there have been indications
of a possible large-scale asymmetry in CMB fluctuations 
\citep{eriksen04, hansen04, bernui06, eriksen07, hansen08,
hoftuft09}. 

Large scale baryon fraction fluctuations would cause such an asymmetry
through differential Thomson scattering. If there are significantly
more electrons along different lines of sight, this leads to a
differential damping of primary CMB fluctuations. The observed dipole
asymmetry in the CMB fluctuations is found to be $0.072 \pm 0.022$
[68\% confidence, \citet{hoftuft09}] in the temperature 
fluctuation amplitude, and the
current best fit for the total Thomson optical depth is $0.087 \pm
0.017$ \citep{dunkley09}. Thus, to explain the observed anomaly would
require variations of $0.8 \pm 0.3$ (at 68\% confidence) in the
baryon fraction on cosmological scales.  Such a large variation (but
only $2.7 \sigma$ from zero) is observed in neither D/H abundances nor
galaxy cluster gas fractions, but smaller variations could be
consistent with both the observed asymmetry and the light-element
constraints.

Thomson scattering by a region with a fluctuating baryon fraction
would also lead to polarization anisotropies in the CMB.  The
polarization anisotropy can be expressed as a sum of ``$E$ modes'' and
``$B$ modes'' \citep{zaldarriaga97}, where $E/B$ refer to curl-free
and curl components of the polarization field.  The primary CMB
components generated by scalar perturbations have zero $B$ modes, with
the dominant primary contributions to $B$ modes sourced by
gravitational radiation.

The CMB polarization signature from an inhomogeneous distribution of
free electrons during and after reionization has two components: that
due to differential screening and that due to rescattering
\citep{dvorkin09}. The differential screening term arises because
modulation of the CMB polarization from recombination by a
direction-dependent $e^{-\tau(\hat{\bf{n}})}$ generates additional $B$
and $E$ modes.  Although the induced $B$ modes from this term are
significant at very large angular scales, they are subdominant at the
intermediate angular scales important for gravitational wave studies.

The other contribution is due to the scattering of incident quadrupole
radiation on free electrons, which generates both $E$- and
$B$-mode polarization if the scatterers are inhomogeneous. On small
scales (where the spherical curvature of the sky can be ignored) the
$B$ modes from rescattering trace the angular fluctuations in the
optical depth on the sky $C_\ell^{\tau\tau}$ as
\citep{hu00,dvorkinsmith09}
\begin{equation}
 C_\ell^{BB {\rm (scatt)}}=C_\ell^{EE {\rm (scatt)}} \simeq \frac{3}{100} Q_{\rm RMS}^2 e^{-2\tau_{\rm eff}} C_\ell^{\tau\tau}
\end{equation}
where $Q_{\rm RMS}$ is the RMS quadrupole, which we take to be 22
$\mu$K, and $\tau_{\rm eff}$ is a characteristic optical depth 
accounting for the low redshift screening of the generated
anisotropies.

\begin{figure}
\plotone{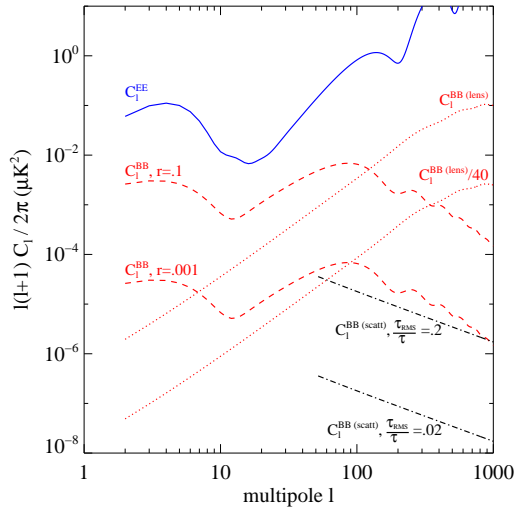}
\caption{The $B$-mode CMB polarization anisotropy induced by scattering
  with a reionized region containing a power-law baryon inhomogeneity
  with total optical depth variations of 20\% (top black dot-dashed)
  and 2\% (bottom black dot-dashed), compared with the $B$-mode signal
  from lensing (top red dotted) and the same curve lowered by a factor
  of 40 as motivated by delensing studies (bottom red
  dotted).  The induced $B$ modes simply trace the assumed optical depth
  fluctuations with a constant prefactor in the flat-sky limit.  The
  red dashed curves show inflation-induced $B$ modes with tensor-to-scalar
  ratios of 0.1 and 0.001.}
\label{fig:rescattered}
\end{figure}

Although a primordial baryon-cdm isocurvature mode could in principle
take any form, with the details set by the physics of how it is
generated, a natural choice is the scale-invariant $k^3 P_B(k) = $
constant, where $P_B(k)=\langle {\delta \rho_B(k) \over \rho_B}
{\delta \rho_B(k)^* \over \rho_B}\rangle$.  The integral along the line of
sight selects a plane of modes in the 3D Fourier space.  This is
simply a slice through the original 3D Fourier space, so the
statistics of the selected modes are unchanged [the ``Limber
  approximation;'' \citet{kaiser92}], and $P_B(k)$ maintains the
same scaling with wavenumber after projection.  Thus we assume
$C_\ell^{\tau\tau} \propto \ell^{-3}$. Figure \ref{fig:rescattered} shows
the rescattered $B$-mode polarization due to this effect, for total
fractional optical depth fluctuations of 20\% and 2\%.  We show
only the result for $\ell > 50$; lower multipoles require a more
sophisticated treatment to include the effects of the spherical
geometry.  

The rescattered $B$ modes in this scenario are well below the $B$
modes induced from $E$ modes through gravitational lensing by
foreground structure.  However, lensing estimation techniques could
reduce the lensing $B$-mode noise by factors of at least 40
\citep{seljak04}.  Rescattered $B$ modes could consequently be a
contaminant in the search for primordial gravitational waves using CMB
polarization, if the tensor-to-scalar ratio is sufficiently low.

\section{Future tests}

With current CMB
data it should be possible to look for variations in the baryon fraction
at recombination. One could either divide the sky into many segments
and analyze each section independently \citep{hansen08}
or use a quadratic estimator [as was done for lensing detection
in \citet{smith07}] to map out baryon fractions on the sphere. 

Large scale structure studies can constrain large scale variations in
the baryon fraction through baryonic effects on the matter transfer
function.  One could use galaxy surveys or QSO counts \citep{hirata09}
to put limits on large scale variations in the matter transfer
function. This will be complicated by the possibility of the baryon
fraction fluctuation being correlated (or anti-correlated) with the
potential fluctuations, but it provides strong constraints. There is
an additional complication in that one needs to understand how the
astrophysics depends on the baryon fraction.  For example, the number
density of high redshift quasars \citep{hirata09} is very sensitive to
the matter power spectrum, but it could also be extremely sensitive to
the baryon fraction.  Cooling processes often scale as density
squared, so formation times could be strongly affected by a varying
baryon density.

Assuming that the varying baryon fraction does not affect the
astrophysics, we can translate the results of \citet{hirata09} into
limits on baryon fraction fluctuations. 
Baryons affect the matter power spectrum
\citep{eisenstein98} such that a 1\% variation in $\sigma_8$ would be 
caused by of order a 5\% change in the baryon fraction.  The constraints of
\citet{hirata09} on large scale gradients in the high redshift
QSO number density (horizon scale gradients in $\ln \sigma_8$
less than 0.03 at 99\%
confidence) are thus comparable to the galaxy cluster gas fraction constraint.

There is significant room for improvement with a larger sample of QSO
absorption studies to constrain intrinsic scatter in D/H.  Deuterium
is an excellent measure of baryon density; with a modest improvement
on current samples it is likely that D/H will be the most robust
constraint on large scale variations in the baryon fraction.

With future radio interferometers such as the Square Kilometer Array,
direct measurements of D/H and $^3$He/H will be possible over a
cosmological volume through radio lines
\citep{sigurdson06,mcquinn09,bagla09}.  The differential Thomson
optical depth can in principle be determined from maps of the neutral
hydrogen density based on redshifted 21cm emission or absorption using
the same interferometers; very large scale variation will be extremely
difficult to separate from foreground emission.

\section{Discussion and conclusion}

We have shown that large scale variation in the cosmological baryon
fraction is constrained, although not strongly, by non-CMB
observations.  Scatter in observed light element abundances constrain
cosmological variations in the baryon density to be less than 26\%
(95\% confidence).  Variations are constrained to be less than 8\%
(95\% confidence) by galaxy cluster measurements, provided there are
no selection effects at work that lead to an artificially low observed
scatter in the gas fraction compared to the cosmic distribution.

Within these bounds of variation, there is an opportunity to ease the
tension between the observed Li and the standard BBN calculation. In
addition, there is evidence of an asymmetry in the large scale CMB
fluctuations 
which could be caused by differential
screening effects.
However, in both cases the observed effects appear to require
fluctuations that are much larger than allowed by galaxy cluster
measurements and well above the upper limits from BBN.  It
is possible that our Galaxy is in an especially baryon-deficient part
of the universe, and that there is an anomalously large baryon
asymmetry on scales of the comoving distance to $z\sim 7$, but this
would require fine-tuning.

Variations in the baryon fraction constitute a mechanism for
generating $B$ modes in CMB polarization; a detection of $B$ modes on
large scales would therefore not be an unambiguous detection of
primordial gravitational waves. The existence of such large scale
baryon fraction modulation would be most simply explained through an
inflation-like mechanism, as there would be super-horizon correlations
in the large scale baryon fraction.

In general, cosmological variations in the baryon fraction (baryon-cdm
isocurvature modes) are allowed at a level that could be important for
galaxy and AGN formation and evolution, and could also be a foreground
for future CMB experiments searching for the signature of primordial 
gravitational radiation. Detection of such a variation would be 
extremely interesting for fundamental physics, as well.

\acknowledgements{We thank the Kavli Insitute for Cosmological
Physics in Chicago, where a significant fraction of this work was done, and
Lloyd Knox, Kendrick Smith, Wayne Hu, and
  Olivier Dore for useful discussions. We acknowledge support
  from the NSERC Discovery Grant program, the Canadian Institute for
  Advanced Research Cosmology \& Gravity program, the Canada
Research Chairs program, and FQRNT.
  K.M.N. is supported by the U. S. Department of Energy, Office of
  Nuclear Physics, under contract No. DE-AC02-06CH11357.  }

\bibliography{baryons}

\begin{thebibliography}{66}
\expandafter\ifx\csname natexlab\endcsname\relax\def\natexlab#1{#1}\fi

\bibitem[{{Allen} {et~al.}(2008){Allen}, {Rapetti}, {Schmidt}, {Ebeling},
  {Morris}, \& {Fabian}}]{allen08}
{Allen}, S.~W., {Rapetti}, D.~A., {Schmidt}, R.~W., {Ebeling}, H., {Morris},
  R.~G., \& {Fabian}, A.~C. 2008, \mnras, 383, 879, 0706.0033

\bibitem[{{Asplund} {et~al.}(2006){Asplund}, {Lambert}, {Nissen}, {Primas}, \&
  {Smith}}]{asplund06}
{Asplund}, M., {Lambert}, D.~L., {Nissen}, P.~E., {Primas}, F., \& {Smith},
  V.~V. 2006, \apj, 644, 229, arXiv:astro-ph/0510636

\bibitem[{{Bagla} \& {Loeb}(2009)}]{bagla09}
{Bagla}, J.~S., \& {Loeb}, A. 2009, ArXiv e-prints, 0905.1698

\bibitem[{{Bania} {et~al.}(2002){Bania}, {Rood}, \& {Balser}}]{bania02}
{Bania}, T.~M., {Rood}, R.~T., \& {Balser}, D.~S. 2002, \nat, 415, 54

\bibitem[{{Bania} {et~al.}(2007){Bania}, {Rood}, \& {Balser}}]{bania07}
------. 2007, Space Science Reviews, 130, 53

\bibitem[{{Bernui} {et~al.}(2006){Bernui}, {Villela}, {Wuensche}, {Leonardi},
  \& {Ferreira}}]{bernui06}
{Bernui}, A., {Villela}, T., {Wuensche}, C.~A., {Leonardi}, R., \& {Ferreira},
  I. 2006, \aap, 454, 409, arXiv:astro-ph/0601593

\bibitem[{{Bonifacio} {et~al.}(2007){Bonifacio}, {Molaro}, {Sivarani},
  {Cayrel}, {Spite}, {Spite}, {Plez}, {Andersen}, {Barbuy}, {Beers}, {Depagne},
  {Hill}, {Fran{\c c}ois}, {Nordstr{\"o}m}, \& {Primas}}]{bonifacio07}
{Bonifacio}, P. {et~al.} 2007, \aap, 462, 851, arXiv:astro-ph/0610245

\bibitem[{{Boyanovsky} {et~al.}(2006){Boyanovsky}, {de Vega}, \&
  {Schwarz}}]{boyanovsky06}
{Boyanovsky}, D., {de Vega}, H.~J., \& {Schwarz}, D.~J. 2006, Annual Review of
  Nuclear and Particle Science, 56, 441, arXiv:hep-ph/0602002

\bibitem[{{Burles} {et~al.}(2001){Burles}, {Nollett}, \& {Turner}}]{burles01}
{Burles}, S., {Nollett}, K.~M., \& {Turner}, M.~S. 2001, \apjl, 552, L1,
  arXiv:astro-ph/0010171

\bibitem[{Burles \& Tytler(1998{\natexlab{a}})}]{burlestytler98a}
Burles, S., \& Tytler, D. 1998{\natexlab{a}}, Astrophys. J., 499, 699

\bibitem[{Burles \& Tytler(1998{\natexlab{b}})}]{burlestytler98b}
------. 1998{\natexlab{b}}, Astrophys. J., 507, 732

\bibitem[{{Charbonnel}(1995)}]{charbonnel95}
{Charbonnel}, C. 1995, \apjl, 453, L41, arXiv:astro-ph/9511080

\bibitem[{{Charbonnel} \& {Primas}(2005)}]{charbonnel05}
{Charbonnel}, C., \& {Primas}, F. 2005, \aap, 442, 961, arXiv:astro-ph/0505247

\bibitem[{{Copi} {et~al.}(1995){Copi}, {Olive}, \& {Schramm}}]{copi95}
{Copi}, C.~J., {Olive}, K.~A., \& {Schramm}, D.~N. 1995, \apj, 451, 51,
  arXiv:astro-ph/9410007

\bibitem[{{Copi} {et~al.}(1998){Copi}, {Olive}, \& {Schramm}}]{copi98}
------. 1998, Proceedings of the National Academy of Science, 95, 2758

\bibitem[{{Crighton} {et~al.}(2004){Crighton}, {Webb}, {Ortiz-Gil}, \&
  {Fern{\'a}ndez-Soto}}]{crighton04}
{Crighton}, N.~H.~M., {Webb}, J.~K., {Ortiz-Gil}, A., \& {Fern{\'a}ndez-Soto},
  A. 2004, \mnras, 355, 1042, arXiv:astro-ph/0403512

\bibitem[{{Deliyannis} {et~al.}(1993){Deliyannis}, {Pinsonneault}, \&
  {Duncan}}]{deliyannis93}
{Deliyannis}, C.~P., {Pinsonneault}, M.~H., \& {Duncan}, D.~K. 1993, \apj, 414,
  740

\bibitem[{{Dunkley} {et~al.}(2009){Dunkley}, {Komatsu}, {Nolta}, {Spergel},
  {Larson}, {Hinshaw}, {Page}, {Bennett}, {Gold}, {Jarosik}, {Weiland},
  {Halpern}, {Hill}, {Kogut}, {Limon}, {Meyer}, {Tucker}, {Wollack}, \&
  {Wright}}]{dunkley09}
{Dunkley}, J. {et~al.} 2009, \apjs, 180, 306, 0803.0586

\bibitem[{{Dvorkin} {et~al.}(2009){Dvorkin}, {Hu}, \& {Smith}}]{dvorkin09}
{Dvorkin}, C., {Hu}, W., \& {Smith}, K.~M. 2009, \prd, 79, 107302, 0902.4413

\bibitem[{{Dvorkin} \& {Smith}(2009)}]{dvorkinsmith09}
{Dvorkin}, C., \& {Smith}, K.~M. 2009, \prd, 79, 043003, 0812.1566

\bibitem[{{Eisenstein} \& {Hu}(1998)}]{eisenstein98}
{Eisenstein}, D.~J., \& {Hu}, W. 1998, \apj, 496, 605, arXiv:astro-ph/9709112

\bibitem[{{Epstein} \& {Petrosian}(1975)}]{epstein75}
{Epstein}, R.~I., \& {Petrosian}, V. 1975, \apj, 197, 281

\bibitem[{{Eriksen} {et~al.}(2007){Eriksen}, {Banday}, {G{\'o}rski}, {Hansen},
  \& {Lilje}}]{eriksen07}
{Eriksen}, H.~K., {Banday}, A.~J., {G{\'o}rski}, K.~M., {Hansen}, F.~K., \&
  {Lilje}, P.~B. 2007, \apjl, 660, L81, arXiv:astro-ph/0701089

\bibitem[{{Eriksen} {et~al.}(2004){Eriksen}, {Hansen}, {Banday}, {G{\'o}rski},
  \& {Lilje}}]{eriksen04}
{Eriksen}, H.~K., {Hansen}, F.~K., {Banday}, A.~J., {G{\'o}rski}, K.~M., \&
  {Lilje}, P.~B. 2004, \apj, 605, 14

\bibitem[{{Geiss} \& {Gloeckler}(1998)}]{geiss98}
{Geiss}, J., \& {Gloeckler}, G. 1998, Space Science Reviews, 84, 239

\bibitem[{{Gloeckler} \& {Geiss}(1996)}]{gloeckler96}
{Gloeckler}, G., \& {Geiss}, J. 1996, \nat, 381, 210

\bibitem[{{Hansen} {et~al.}(2004){Hansen}, {Banday}, \&
  {G{\'o}rski}}]{hansen04}
{Hansen}, F.~K., {Banday}, A.~J., \& {G{\'o}rski}, K.~M. 2004, \mnras, 354,
  641, arXiv:astro-ph/0404206

\bibitem[{{Hansen} {et~al.}(2008){Hansen}, {Banday}, {Gorski}, {Eriksen}, \&
  {Lilje}}]{hansen08}
{Hansen}, F.~K., {Banday}, A.~J., {Gorski}, K.~M., {Eriksen}, H.~K., \&
  {Lilje}, P.~B. 2008, ArXiv e-prints, 0812.3795

\bibitem[{{Hirata}(2009)}]{hirata09}
{Hirata}, C.~M. 2009, ArXiv e-prints, 0907.0703

\bibitem[{{Hoftuft} {et~al.}(2009){Hoftuft}, {Eriksen}, {Banday}, {G{\'o}rski},
  {Hansen}, \& {Lilje}}]{hoftuft09}
{Hoftuft}, J., {Eriksen}, H.~K., {Banday}, A.~J., {G{\'o}rski}, K.~M.,
  {Hansen}, F.~K., \& {Lilje}, P.~B. 2009, \apj, 699, 985, 0903.1229

\bibitem[{{Hogan}(1995)}]{hogan95}
{Hogan}, C.~J. 1995, \apjl, 441, L17, arXiv:astro-ph/9407038

\bibitem[{{Hu}(2000)}]{hu00}
{Hu}, W. 2000, \apj, 529, 12, arXiv:astro-ph/9907103

\bibitem[{{Izotov} {et~al.}(2007){Izotov}, {Thuan}, \&
  {Stasi{\'n}ska}}]{izotov07}
{Izotov}, Y.~I., {Thuan}, T.~X., \& {Stasi{\'n}ska}, G. 2007, \apj, 662, 15,
  arXiv:astro-ph/0702072

\bibitem[{{Jedamzik} \& {Fuller}(1995)}]{jedamzik95}
{Jedamzik}, K., \& {Fuller}, G.~M. 1995, \apj, 452, 33, arXiv:astro-ph/9410027

\bibitem[{{Kaiser}(1992)}]{kaiser92}
{Kaiser}, N. 1992, \apj, 388, 272

\bibitem[{{Kirkman} {et~al.}(2000){Kirkman}, {Tytler}, {Burles}, {Lubin}, \&
  {O'Meara}}]{kirkman00}
{Kirkman}, D., {Tytler}, D., {Burles}, S., {Lubin}, D., \& {O'Meara}, J.~M.
  2000, \apj, 529, 655, arXiv:astro-ph/9907128

\bibitem[{{Kirkman} {et~al.}(2003){Kirkman}, {Tytler}, {Suzuki}, {O'Meara}, \&
  {Lubin}}]{kirkman03}
{Kirkman}, D., {Tytler}, D., {Suzuki}, N., {O'Meara}, J.~M., \& {Lubin}, D.
  2003, \apjs, 149, 1, arXiv:astro-ph/0302006

\bibitem[{{Kurki-Suonio} {et~al.}(1997){Kurki-Suonio}, {Jedamzik}, \&
  {Mathews}}]{kurki97}
{Kurki-Suonio}, H., {Jedamzik}, K., \& {Mathews}, G.~J. 1997, \apj, 479, 31,
  arXiv:astro-ph/9606011

\bibitem[{{Lara} {et~al.}(2006){Lara}, {Kajino}, \& {Mathews}}]{lara06}
{Lara}, J.~F., {Kajino}, T., \& {Mathews}, G.~J. 2006, \prd, 73, 083501,
  arXiv:astro-ph/0603817

\bibitem[{{Levshakov} {et~al.}(2002){Levshakov}, {Dessauges-Zavadsky},
  {D'Odorico}, \& {Molaro}}]{levskakov02}
{Levshakov}, S.~A., {Dessauges-Zavadsky}, M., {D'Odorico}, S., \& {Molaro}, P.
  2002, \apj, 565, 696, arXiv:astro-ph/0105529

\bibitem[{{Malaney} \& {Mathews}(1993)}]{malaney93}
{Malaney}, R.~A., \& {Mathews}, G.~J. 1993, \physrep, 229, 145

\bibitem[{{McQuinn} \& {Switzer}(2009)}]{mcquinn09}
{McQuinn}, M., \& {Switzer}, E.~R. 2009, ArXiv e-prints, 0905.1715

\bibitem[{{Nagai} {et~al.}(2007){Nagai}, {Vikhlinin}, \& {Kravtsov}}]{nagai07}
{Nagai}, D., {Vikhlinin}, A., \& {Kravtsov}, A.~V. 2007, \apj, 655, 98,
  arXiv:astro-ph/0609247

\bibitem[{{Olive} \& {Skillman}(2004)}]{olive04}
{Olive}, K.~A., \& {Skillman}, E.~D. 2004, \apj, 617, 29,
  arXiv:astro-ph/0405588

\bibitem[{{Olive} {et~al.}(1997){Olive}, {Steigman}, \& {Skillman}}]{olive97}
{Olive}, K.~A., {Steigman}, G., \& {Skillman}, E.~D. 1997, \apj, 483, 788,
  arXiv:astro-ph/9611166

\bibitem[{{O'Meara} {et~al.}(2006){O'Meara}, {Burles}, {Prochaska}, {Prochter},
  {Bernstein}, \& {Burgess}}]{omeara06}
{O'Meara}, J.~M., {Burles}, S., {Prochaska}, J.~X., {Prochter}, G.~E.,
  {Bernstein}, R.~A., \& {Burgess}, K.~M. 2006, \apjl, 649, L61,
  arXiv:astro-ph/0608302

\bibitem[{{O'Meara} {et~al.}(2001){O'Meara}, {Tytler}, {Kirkman}, {Suzuki},
  {Prochaska}, {Lubin}, \& {Wolfe}}]{omeara01}
{O'Meara}, J.~M., {Tytler}, D., {Kirkman}, D., {Suzuki}, N., {Prochaska},
  J.~X., {Lubin}, D., \& {Wolfe}, A.~M. 2001, \apj, 552, 718,
  arXiv:astro-ph/0011179

\bibitem[{{Peimbert} {et~al.}(2007){Peimbert}, {Luridiana}, \&
  {Peimbert}}]{peimbert07}
{Peimbert}, M., {Luridiana}, V., \& {Peimbert}, A. 2007, \apj, 666, 636,
  arXiv:astro-ph/0701580

\bibitem[{{Peimbert} {et~al.}(2000){Peimbert}, {Peimbert}, \&
  {Ruiz}}]{peimbert00}
{Peimbert}, M., {Peimbert}, A., \& {Ruiz}, M.~T. 2000, \apj, 541, 688,
  arXiv:astro-ph/0003154

\bibitem[{{Pettini} \& {Bowen}(2001)}]{pettini01}
{Pettini}, M., \& {Bowen}, D.~V. 2001, \apj, 560, 41, arXiv:astro-ph/0104474

\bibitem[{{Pettini} {et~al.}(2008){Pettini}, {Zych}, {Murphy}, {Lewis}, \&
  {Steidel}}]{pettini08}
{Pettini}, M., {Zych}, B.~J., {Murphy}, M.~T., {Lewis}, A., \& {Steidel}, C.~C.
  2008, \mnras, 391, 1499, 0805.0594

\bibitem[{{Ryan} {et~al.}(1999){Ryan}, {Norris}, \& {Beers}}]{ryan99}
{Ryan}, S.~G., {Norris}, J.~E., \& {Beers}, T.~C. 1999, \apj, 523, 654,
  arXiv:astro-ph/9903059

\bibitem[{{Schramm} \& {Turner}(1998)}]{schramm98}
{Schramm}, D.~N., \& {Turner}, M.~S. 1998, Reviews of Modern Physics, 70, 303,
  arXiv:astro-ph/9706069

\bibitem[{{Seljak} \& {Hirata}(2004)}]{seljak04}
{Seljak}, U., \& {Hirata}, C.~M. 2004, \prd, 69, 043005, arXiv:astro-ph/0310163

\bibitem[{{Sigurdson} \& {Furlanetto}(2006)}]{sigurdson06}
{Sigurdson}, K., \& {Furlanetto}, S.~R. 2006, Physical Review Letters, 97,
  091301, arXiv:astro-ph/0505173

\bibitem[{{Smith} {et~al.}(2007){Smith}, {Zahn}, \& {Dor{\'e}}}]{smith07}
{Smith}, K.~M., {Zahn}, O., \& {Dor{\'e}}, O. 2007, \prd, 76, 043510, 0705.3980

\bibitem[{{Sollom} {et~al.}(2009){Sollom}, {Challinor}, \& {Hobson}}]{sollom09}
{Sollom}, I., {Challinor}, A., \& {Hobson}, M.~P. 2009, \prd, 79, 123521,
  0903.5257

\bibitem[{{Spite} \& {Spite}(1982)}]{spite82}
{Spite}, F., \& {Spite}, M. 1982, \aap, 115, 357

\bibitem[{{Thorburn}(1994)}]{thorburn94}
{Thorburn}, J.~A. 1994, \apj, 421, 318

\bibitem[{{Tytler} {et~al.}(1996){Tytler}, {Fan}, \& {Burles}}]{tytler96}
{Tytler}, D., {Fan}, X.-M., \& {Burles}, S. 1996, \nat, 381, 207,
  arXiv:astro-ph/9603069

\bibitem[{{Vikhlinin} {et~al.}(2006){Vikhlinin}, {Kravtsov}, {Forman}, {Jones},
  {Markevitch}, {Murray}, \& {Van Speybroeck}}]{vikhlinin06}
{Vikhlinin}, A., {Kravtsov}, A., {Forman}, W., {Jones}, C., {Markevitch}, M.,
  {Murray}, S.~S., \& {Van Speybroeck}, L. 2006, \apj, 640, 691,
  arXiv:astro-ph/0507092

\bibitem[{{Wagoner}(1973)}]{wagoner73}
{Wagoner}, R.~V. 1973, \apj, 179, 343

\bibitem[{{Wasserburg} {et~al.}(1995){Wasserburg}, {Boothroyd}, \&
  {Sackmann}}]{wasserburg95}
{Wasserburg}, G.~J., {Boothroyd}, A.~I., \& {Sackmann}, I.-J. 1995, \apjl, 447,
  L37

\bibitem[{{White} {et~al.}(1993){White}, {Navarro}, {Evrard}, \&
  {Frenk}}]{white93}
{White}, S.~D.~M., {Navarro}, J.~F., {Evrard}, A.~E., \& {Frenk}, C.~S. 1993,
  \nat, 366, 429

\bibitem[{{Witten}(1984)}]{witten84}
{Witten}, E. 1984, \prd, 30, 272

\bibitem[{{Zaldarriaga} \& {Seljak}(1997)}]{zaldarriaga97}
{Zaldarriaga}, M., \& {Seljak}, U. 1997, \prd, 55, 1830, arXiv:astro-ph/9609170

\end{thebibliography}
\bibliographystyle{hapj}

\end{document}